\documentclass[%
reprint,
superscriptaddress,
footinbib,
twocolumn,
amsmath,amssymb,
aps,
floatfix,
]{revtex4-1}

\usepackage{graphicx}
\usepackage{dcolumn}
\usepackage{bm}
\usepackage{amssymb}
\usepackage{multirow}
\usepackage[colorlinks=true,citecolor=blue]{hyperref}

\begin{document}

\title{Berezinskii-Kosterlitz-Thouless Phase in Two-dimensional Ferroelectrics}

\author{Changsong Xu}
\altaffiliation{Contributed equally to this work}
\affiliation{Physics Department and Institute for Nanoscience and Engineering, University of Arkansas, Fayetteville, Arkansas 72701, USA}%

\author{Yousra Nahas}
\altaffiliation{Contributed equally to this work}
\affiliation{Physics Department and Institute for Nanoscience and Engineering, University of Arkansas, Fayetteville, Arkansas 72701, USA}%

\author{Sergei Prokhorenko}
\affiliation{Physics Department and Institute for Nanoscience and Engineering, University of Arkansas, Fayetteville, Arkansas 72701, USA}%

\author{Hongjun Xiang}
\email{hxiang@fudan.edu.cn}
\affiliation{Key Laboratory of Computational Physical Sciences (Ministry of Education), State Key Laboratory of Surface Physics, and Department of Physics, Fudan University, Shanghai, 200433, China}%
\affiliation{Collaborative Innovation Center of Advanced Microstructures, Nanjing 210093, China}

\author{L. Bellaiche}
\email{laurent@uark.edu}
\affiliation{Physics Department and Institute for Nanoscience and Engineering, University of Arkansas, Fayetteville, Arkansas 72701, USA}%


\begin{abstract}
  The celebrated Berezinskii-Kosterlitz-Thouless (BKT) phase transition refers to a topological transition characterized, e.g., by the dissociation of vortex-antivortex pairs in two-dimensional (2D) systems.
  Such unusual phase has been reported in various types of materials, but never in the new class of systems made by {\it one-unit cell-thick} (1UC) ferroelectrics (also coined as 2D ferroelectrics).   Here, the use of  a first-principles-based effective Hamiltonian method leads to the discovery  of many fingerprints of a BKT phase existing in-between the ferroelectric and paraelectric states of 1UC tin tellurium being fully relaxed. Moreover, epitaxial strain is found to have dramatic consequences on the temperature range of such BKT phase for the 1UC SnTe.
  Consequently, our predictions  extend the playground of BKT theory to a novel class of functional materials, and demonstrate that strain is an effective tool to alter BKT characteristics there.
 \end{abstract}


\maketitle

According to the Mermin-Wagner's theorem, spontaneous symmetry breaking is prohibited in two-dimensional (2D) systems with continuous symmetry and short-range interactions, as long-range orders are suppressed by the strong fluctuations \cite{mermin1966absence}.
However, such theorem does not prevent the transition to a topological state populated by neutral bound pairs of vortex and antivortex, which is known as the Berezinskii-Kosterlitz-Thouless (BKT) phase transition \cite{berezinsky1972destruction,kosterlitz1973ordering,kosterlitz1974critical}.
Such BKT phase transition is of infinite order, with its physics being captured/approximated by the 2D XY model \cite{berezinsky1972destruction,kosterlitz1973ordering,kosterlitz1974critical}.
The BKT phase is characterized by a quasi-long-range order for which the correlation function obeys a power law. Note that BKT phases have been observed in various types of systems, including Van der Waals layers \cite{johansen2019current}, films made of superfluid helium  \cite{bishop1978study} and of superconductors \cite{beasley1979possibility,hebard1980evidence,wolf1981two}, 2D electron gas \cite{frohlich1981kosterlitz}, Josephson junction arrays \cite{resnick1981kosterlitz} and nematic liquid crystals \cite{lammert1993topology}.

\begin{figure}[b]
\centering
  \includegraphics[width=8cm]{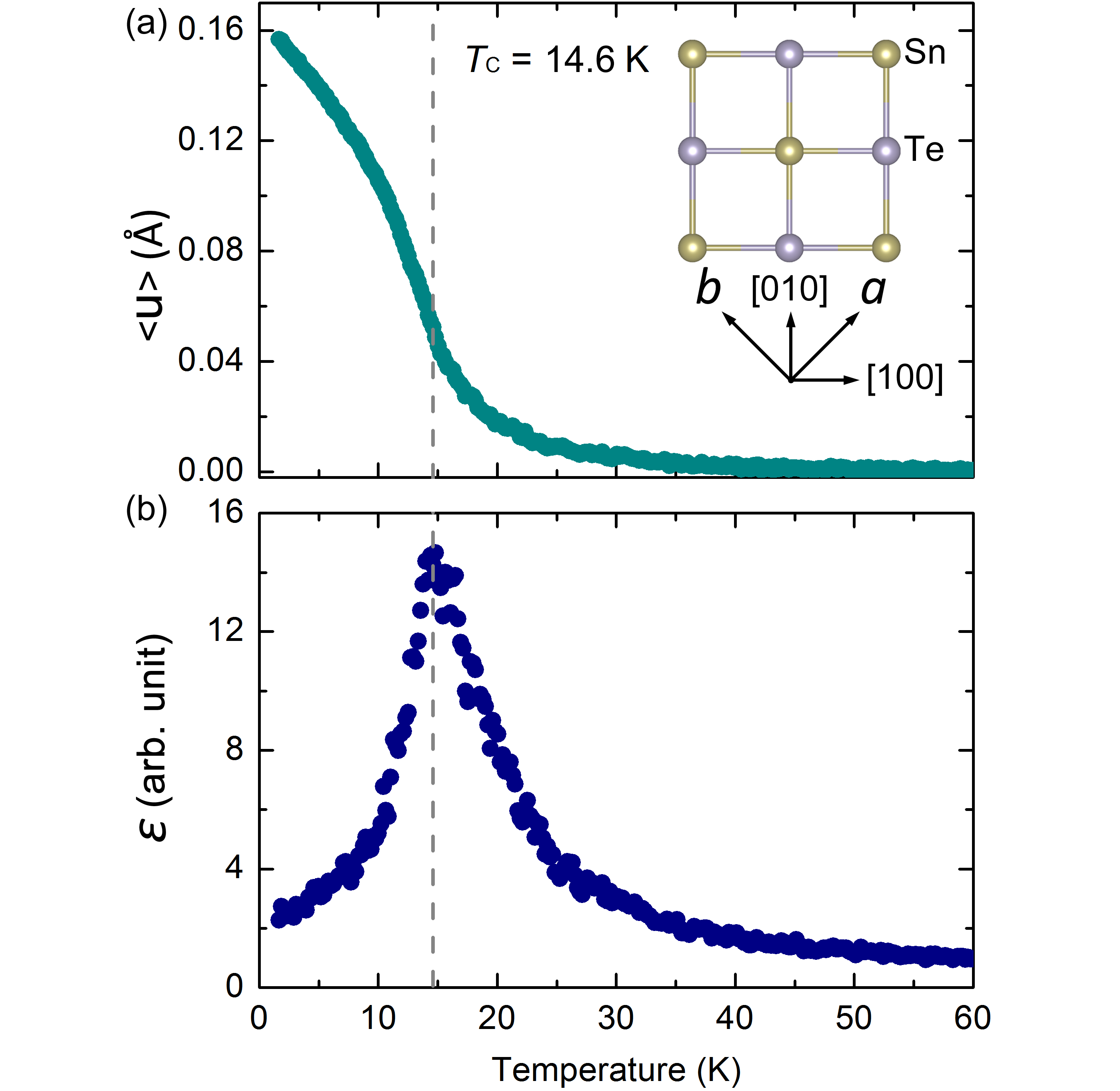}%
  \caption{ Predicted macroscopic properties of relaxed 1UC SnTe  film as a function of temperature. Panel (a) displays the supercell average of the local modes, with its inset showing the top view of SnTe structure and its axes; Panel (b) shows the dielectric response.}
\end{figure}

\begin{figure}[t]
\centering
  \includegraphics[width=8cm]{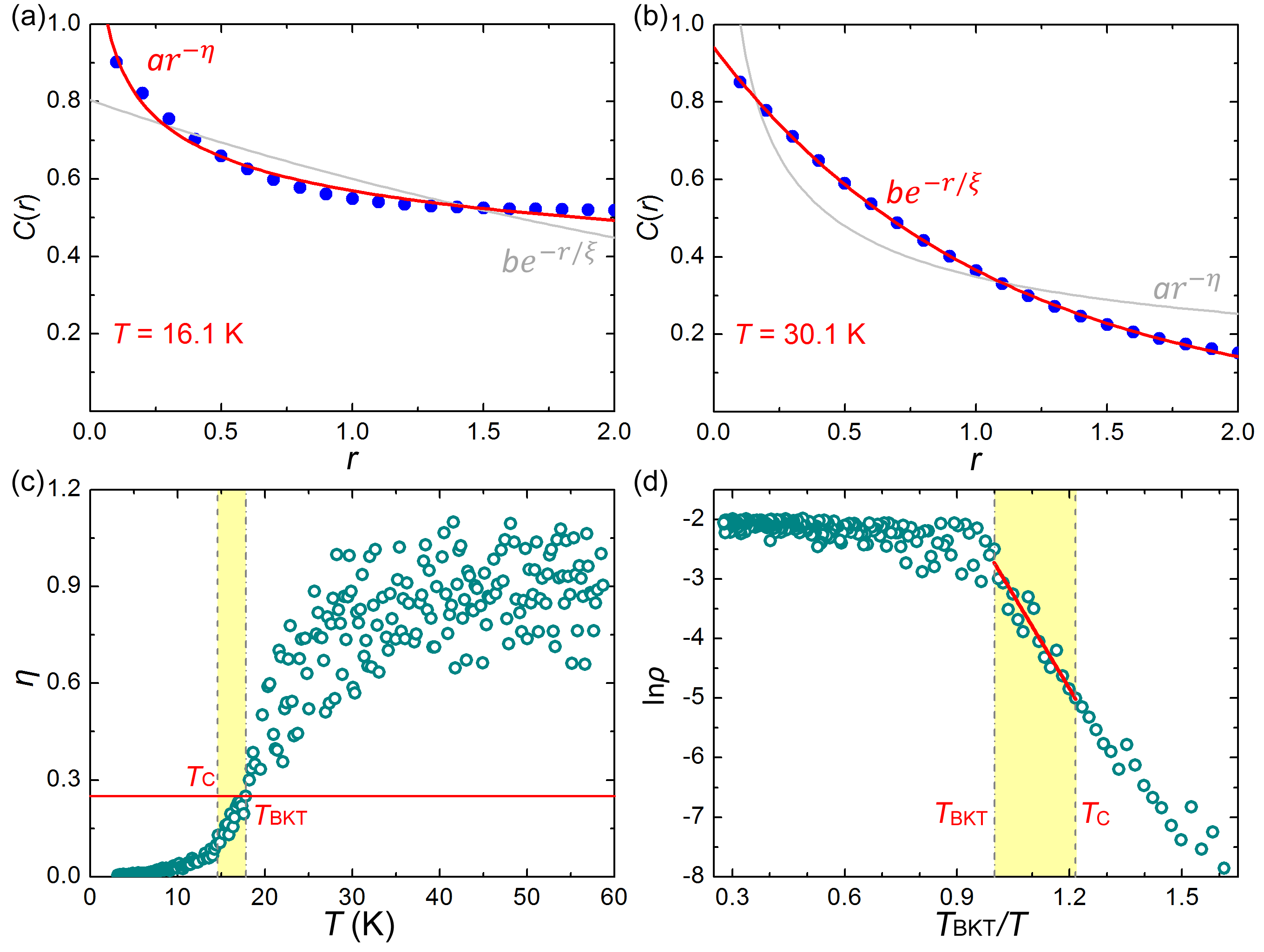}%
  \caption{Predicted local ordering of relaxed 1UC SnTe film. Panels (a) and (b) show the spatial correlations at 16.1 and 30.1K, respectively, with fits by appropriate functions being reported via solid red lines;  Panel (c) reports  the evolution of the corresponding correlation function exponent $\eta$ with temperature. Panel (d) displays the evolution of the natural logarithm  of the average vortex-antivortex pair density with the ratio of $T/T_{\rm BKT}$, where $T$ is the temperature and $T_{\rm BKT}$ is the critical temperature at which the BKT phase transforms into the disordered state.}
\end{figure}

Interestingly,  the long-range dipolar interactions inherent to ferroelectrics \cite{cohen1992origin}, and their effect on  strongly suppressing fluctuations,  were naturally thought to prevent the occurrence of a BKT phase in ferroelectric ultrathin films. However, a recent work has predicted the existence of such BKT phase for some temperature region, for which continuous symmetry is basically satisfied and bridging the ferroelectric (FE) and paraelectric (PE) phases,   in BaTiO$_3$ thin films being three monolayer-thick  and being subjected to a 3\% tensile epitaxial train \cite{nahas2017emergent}. Such prediction opens the door to the search for a BKT phase in the new class of materials made by 2D ferroelectrics and that consists of {\it one-unit cell-thick} (1UC) ferroelectrics \cite{chang2016discovery,fei2016ferroelectricity,zhou2017out,zheng2018room}. This new class of materials differentiates itself from ultrathin films of perovskites because they can still exhibit ferroelectricity for the ultimate thickness of one unit-cell  while long-range-ordering of dipoles disappears in ultrathin films of perovskites (such as BaTiO$_3$) typically below 3-6 unit cells (see, e.g., Refs. \cite{lai2007thickness,junquera2003critical} and references therein).
For instance, the topological crystalline insulator SnTe has been recently reported to possess an in-plane polarization down to 1UC in thickness \cite{chang2016discovery}, which motivates us to investigate the BKT phase in such promising 2D system.
Furthermore, one may wonder how epitaxial strain (which is an handle that has been demonstrated to drastically change properties of perovskites, see, e.g., Refs. \cite{zeches2009strain,chen2015large,xu2017novel,tan2016pressure} and references therein) can affect the existence and stability of a BKT phase (if any) in these 2D ferroelectrics.

In this Letter, we basically follow the procedure used in Ref. \cite{nahas2017emergent} and take advantage of a newly developed first-principle-based effective Hamiltonian method combined with Monte Carlo (MC) simulations, in order to  investigate the 2D ferroelectric SnTe material, in general, and report (i) the prediction of many features consistent with a BKT phase in 1UC SnTe and (ii) how strain affects the temperature region at which the BKT phase is stable.

\begin{figure*}[t]
\centering
  \includegraphics[width=16cm]{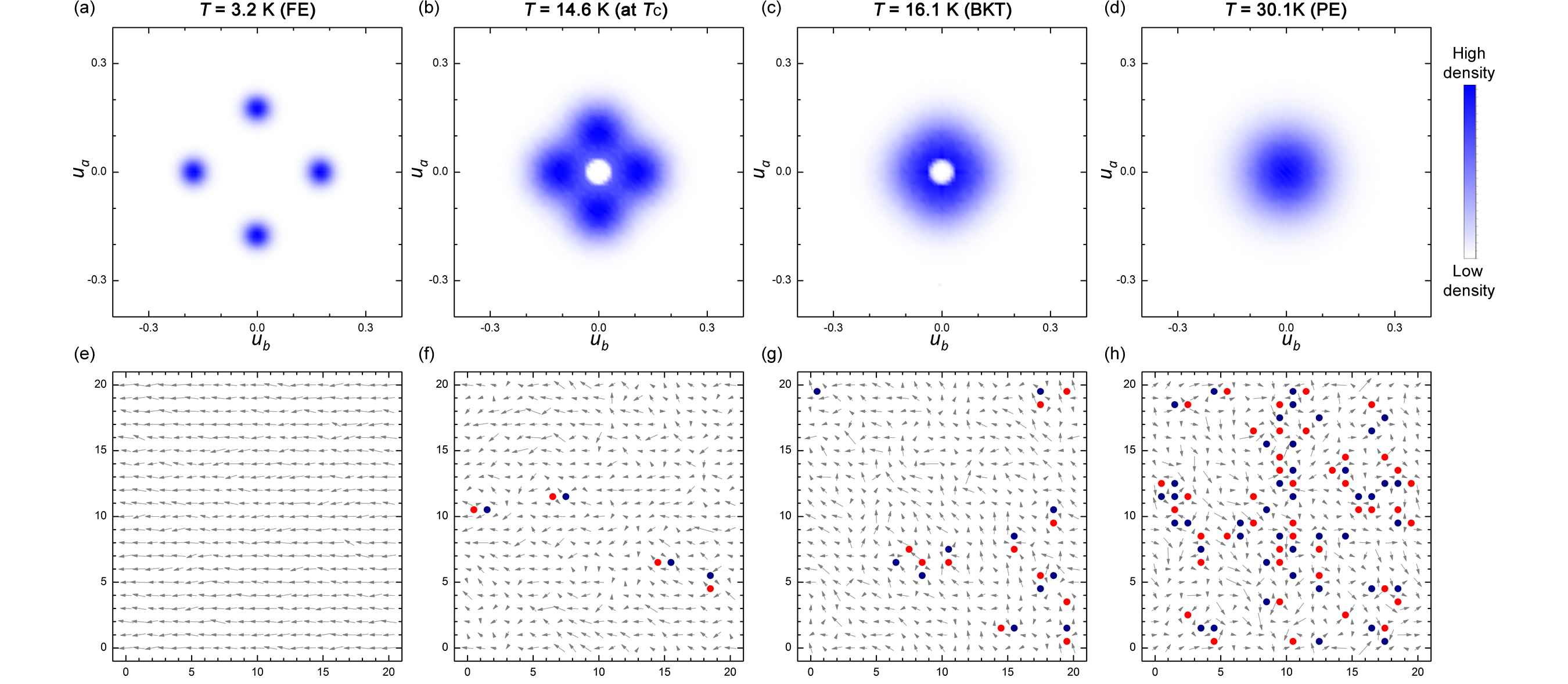}%
  \caption{Local  properties of relaxed 1UC SnTe film. Panels (a)-(d) report the symmetrized probability distribution of local modes for four different temperatures. Panels (e)-(h) display
  snapshots of dipolar configurations within a plane spanned by the [110] and [1$\bar{1}$0] directions, for these four temperatures, with vortices and antivortices being indicated in blue and red colors, respectively. 
  }
  \end{figure*}

Here, we use the effective Hamiltonian scheme that was recently developed and used to study SnTe films \cite{liu2018intrinsic}. Its degree of freedoms are: (1) the  $\bm{u_i}$  local soft mode that corresponds to the relative polar displacements of Sn ions at site $i$;
and (2) the ${\epsilon_l}$ strain tensor that contains both homogeneous and inhomogeneous parts. Note that in the present case of thin films, only the three in-plane strain components ($\epsilon_l=\epsilon_1,\epsilon_2$ and $\epsilon_6$ in Voigt notation) are considered to be variables. This Hamiltonian possesses the following four energetic terms:
\begin{equation}\label{}
\begin{aligned}
  E^{tot}=&E^{self}(\{\bm{u_i}\}) + E^{pair}(\{\bm{u_i}\})  + E^{elas}(\{\bm{\epsilon_l}\}) \\
          &+ E^{int}(\{\bm{u_i}\},\{\bm{\epsilon_l}\}),
\end{aligned}
\end{equation}
where $E^{self}$ is the local-mode self-energy, for which up to fourth-order expansions in $\bm{u_i}$ are included; $E^{pair}(\{\bm{u}\}) = \frac{1}{2} \sum_{i \neq j} \sum_{\alpha\beta} J_{ij,\alpha\beta} u_{i\alpha} u_{j\beta}$ gathers {\it all} the bilinear interactions between Sn-Sn pairs being distant by less than 12.9 \AA . In other words, $E^{pair}$ contains both short-range and long-range dipolar interactions up to the ninth Sn-Sn nearest neighbors (note that such treatment naturally allows to implicitly take into account the significant spatial dependence of the electronic dielectric constant, that is involved in the long-range dipolar interactions \cite{zhong1995first,bellaiche2000finite}, known to occur in 2D systems \cite{qiu2016screening});  an elastic energy, $E^{elas}(\{\bm{\epsilon_l}\})$; and a coupling between strain and local modes of the form $E^{int}(\{\bm{u_i}\},\{\bm{\epsilon_l}\}) = \frac{1}{2} \sum_i \sum_{l\alpha\beta} B_{l\alpha\beta} \bm{\epsilon_l} u_{\alpha}(\bm{R}_i) u_{\beta}(\bm{R}_i)$.
The analytical expressions of $E^{self}$, $E^{elas}$ and $E^{int}$ are identical to the corresponding  ones appearing in the effective Hamiltonians for  ferroelectric perovskite oxides \cite{zhong1995first,bellaiche2000finite}. All the parameters of the effective Hamiltonian are computed by performing first-principles calculations on small supercells. For instance, the elastic-local mode interacting parameters, $B_{l\alpha\beta}$, are obtained by considering the dependence of the force constants on strain, as predicted by Density Functional Theory (DFT). Note also that the effective Hamiltonian parameters are thickness-dependent, that is they change when varying the thickness -- which is in-line with the non-monotonic behavior of the Curie temperature with thickness that was experimentally found in SnTe films \cite{chang2016discovery}.

Once these parameters are determined, we employ them, along with the resulting effective Hamiltonian, on 20$\times$20$\times$$n$ supercells with thickness $n$ varying from 1UC to 3UC. Such supercells are periodic along the in-plane directions, while having vacuum layers along the out-of-plane direction. The effect of the substrates on properties is simply mimicked by imposing an epitaxial constraint on the in-plane lattice vectors of the SnTe systems. Technically,  we performed parallel tempering Monte Carlo (PTMC) simulations \cite{wang2015predicting,hukushima1996exchange} using heat bath algorithm \cite{miyatake1986implementation}, in order to obtain finite-temperature properties of SnTe thin films.

Let us mostly concentrate on a single monolayer made of SnTe that is fully relaxed (i.e., for which the strain elements $\epsilon_1,\epsilon_2$ and $\epsilon_6$ adopt values minimizing the total energy).
The strains are computed with respect to $a=b=$ 4.553 \AA, which are the lattice constants of 1UC SnTe in the PE state from DFT. The optimized lattice constants for FE state from Eq. (1) yields $a$ = 4.548 \AA~and $b$ = 4.568 \AA, which is consistent with the measured 4.44 $\pm$ 0.05 \AA~and 4.58 $\pm$ 0.05 \AA, respectively \cite{chang2016discovery}.
Figure 1a displays the evolution of its total local mode (averaged over the sites), $<$${\bm u}$$>$, with temperature, $T$. Such averaged local mode is
nearly zero at high temperature, which   corresponds to a disordered paraelectric (PE) phase of tetragonal $P4/nmm$ symmetry.  On the other hand, at lower temperature, the supercell average of the local mode is finite and lying along the [110] direction, reflecting the existence of an electrical polarization being parallel to the (in-plane) [110] direction and thus to the occurrence of a polar phase of orthorhombic $P2_1mn$ ($Pmn2_1$ in standard setting) symmetry. The predicted  Curie transition temperature can be first obtained by determining the inflection point of the $<$${\bm u}$$>$-{\it versus}-$T$ curve, which provides a value of $T_{\rm C}$ = 14.6 K.
Such latter value is confirmed by computing the  dielectric constant and determining its maximum with temperature, as revealed by Figure 1b. Note that a paraelectric-to-ferroelectric transition has indeed been observed in SnTe monolayer \cite{chang2016discovery}, but with a $T_{\rm C}$ of 270 K. Such discrepancy may reflect a typical underestimation of the Curie temperature by effective Hamiltonian schemes \cite{zhong1995first,ye2018ferroelectric} but also the possibility that the measured $T_{\rm C}$ owes its large value to the existence of structural defects in the grown sample that may, e.g. affect strain around them (we will discuss later on how $T_{\rm C}$ can be dramatically affected by strain).

Let us now try to predict if a BKT phase can exist  in relaxed 1UC SnTe. For that, we computed the disconnected correlation function\cite{newman1999monte}, $C(r) \equiv$ $<$$\bm{u_i} \cdot \bm{u_j}$$>$, as a function of the distance $r$  separating the  $i$ and $j$ sites, for different temperatures. As demonstrated in Figs. 2a and 2b, this correlation function is well described by an exponential decrease of the form $be^{-r/\xi}$ for high temperatures in the disordered PE phase, while it follows rather closely a power law decay $ar^{-\eta}$ with distance for some low temperatures.
Such difference in behavior is precisely what is expected when going from a ``typical'' state to a BKT phase when cooling the system under investigation \cite{kosterlitz1973ordering,binder1981finite,binder1981k,binder1985finite,landau2014guide,baumgartner2013applications}. In order to further assure that we do have a
BKT phase in the studied system, as well as to precisely determine the upper critical temperature at which such phase occurs, Figure 2c reports the temperature evolution of the  $\eta$ parameter involved in the aforementioned power law decay. This parameter becomes equal to 0.25 at the temperature of 17.8K, which is exactly the critical value that is expected when a BKT phase emerges \cite{berezinsky1972destruction,kosterlitz1973ordering}. Considering such fact, as well as the existence of a Curie temperature at about 14.6K, one can then deduce that the present calculations predict the existence of a BKT phase within the narrow range of 14.6--17.8K and  that bridges the PE to FE phases.  Such intermediate nature of a  BKT phase has also been recently numerically found in thin films made of BaTiO$_3$ perovskite \cite{nahas2017emergent}.

Furthermore, Figures 3a-d display the symmetrized probability distribution of local modes \cite{prokhorenko2017fluctuations} for four different temperatures: 3.2 K, which corresponds to the ferroelectric phase, 14.6 and 16.1K that both lie within the BKT phase region, and 30.1K that is within the disordered PE phase. Four distinct and relative small spots can be seen at the lowest temperature in Fig. 3a, as consistent with a fourfold-degenerate ferroelectric ground state of $P2_1mn$ symmetry. Note that the polarization in the ferroelectric phase has an equal probability to  lie along one of the four [110], [$\bar{1}$10], [1$\bar{1}$0] and [$\bar{1}$$\bar{1}$0] directions, but, once the systems picks up one of these four possibilities, the polarization stays along that direction rather than fluctuates among other $<$110$>$ minima. On the other hand, at the highest temperature, a radial symmetric  distribution centered around zero values occurs in Fig. 3d, which reflects the disordered character of the PE state. Interestingly, the transition from the symmetrized probability distribution of local modes of Fig. 3a to that of Fig. 3d happens under heating within the BKT phase, by first enlarging
the four isolated spots (which indicates a larger fluctuation of the electric dipoles, see Fig. 3b) until they merge together ({\it cf} Fig. 3c) creating a more continuous, but still approximate, rotational symmetry.

\begin{figure}[t]
\centering
  \includegraphics[width=8cm]{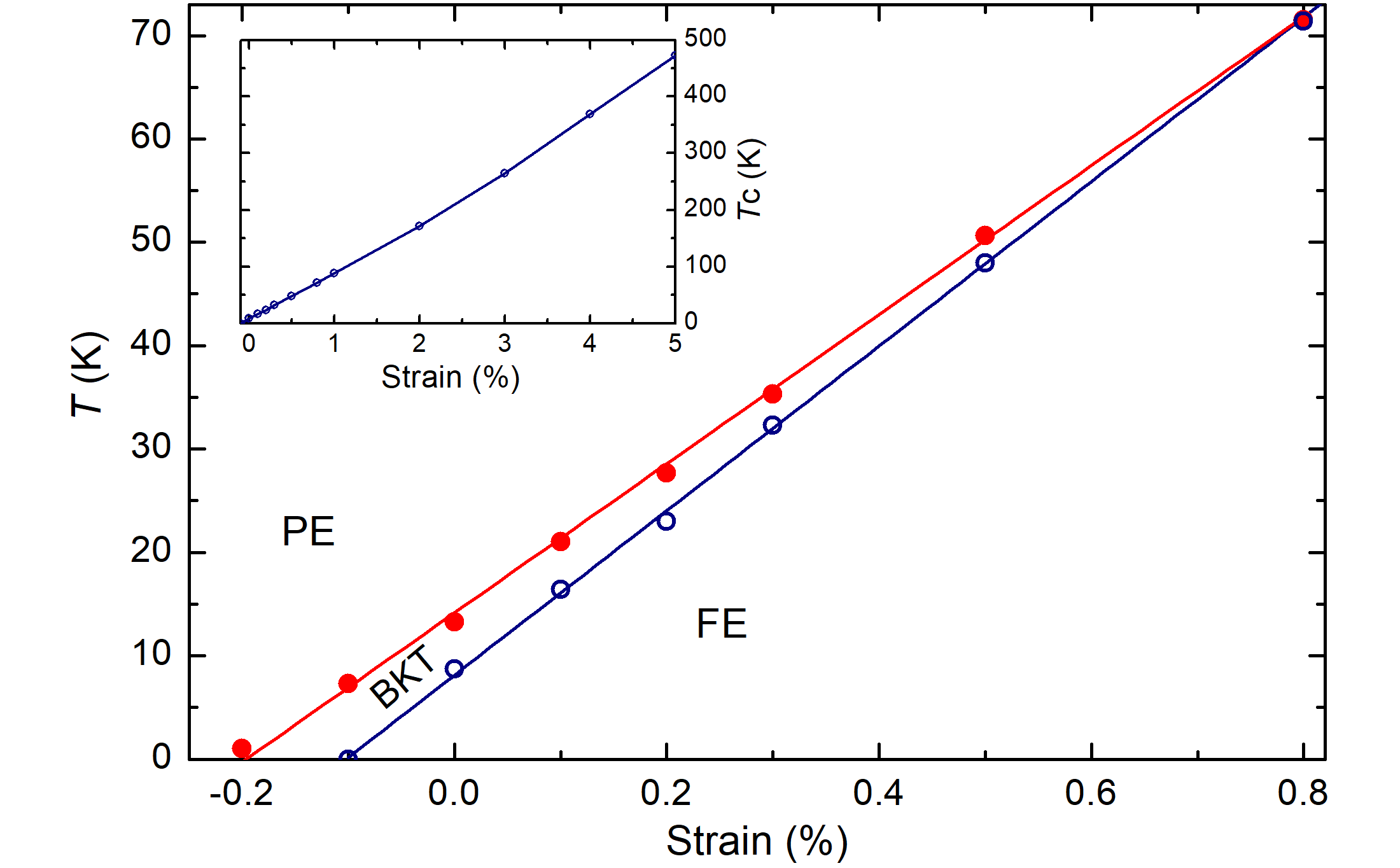}%
  \caption{Temperature-{\it vs}-strain phase diagram of 1UC SnTe. The horizontal axis represents the epitaxial strain $\epsilon_{misfit} = \epsilon_1 = \epsilon_2$ (while $\epsilon_6$ = 0). $T_{\rm BKT}$ and $T_{\rm C}$ are shown by red dots and blue circles, respectively, from which the red and blue lines are fitted, respectively. The inset displays $T_{\rm C}$ for a wider strain range.}
  \end{figure}

Moreover, it is well-known that a BKT phase involves the dynamical pairing between vortices and antivortices \cite{kosterlitz1973ordering,tobochnik1979monte}. In order to further confirm our provocative prediction of a BKT state in a relaxed SnTe monolayer, we therefore decided to investigate and report in Figs. 3e-h dipolar patterns obtained at the different temperatures associated with Figs 3a-d. Figures 3e shows that, for temperatures below $T_{\rm C}$, the local modes are mostly all parallel to each other, with no vortices and antivortices occurring -- as consistent with an ``usual'' ferroelectric state. On the other hand, when temperature increases and crosses $T_{\rm C}$ (see Figure 3f), tightly bound electrical vortex-antivortex pairs do emerge, which is, once again, in-line with our  finding of a BKT transition in the presently studied 2D ferroelectric system. As temperature further increases but still within the BKT phase (that is from 14.6 to 17.8K), Fig. 3g demonstrates that vortices and antivortices are more numerous but kept bound to each other.  Finally, for temperature above $T_{\rm BKT}=17.8K$ ({\it cf} Fig. 3h), these pairs appear to be unbound while their density keep increasing. All these findings therefore confirm the prediction of a BKT phase in relaxed 1UC SnTe system, within a narrow temperature range above $T_{\rm C}$. Note that the Supplementary Material (SM) \cite{sm} provides additional information demonstrating that our presently discovered BKT phase is dynamical in nature, as well as possesses correlation lengths that diverge around $T_{\rm BKT}$, as expected \cite{kosterlitz1973ordering}.

Figure 2d further reports the natural logarithm  of the $\rho$ density of vortex-antivortex pairs as a function of the inverse of the temperature. In addition to numerically confirm that the highest number of topological defects occurs at high temperature (as consistent with Fig. 3h), Figure 2d also reveals that  $ln(\rho)$ adopts a linear behavior with  the ratio $T_{\rm BKT}/T$ with a slope being given by -10.52 $\pm$ 0.05 within the predicted BKT phase, which agrees very well with the value of of -10.2 that is predicted to create a closely bound vortex-antivortex pair
in the BKT phase \cite{kosterlitz1973ordering,kosterlitz1974critical}.

Note that we also checked the dependency of the lateral periodic lengths on the results for the 1UC film, by  performing additional calculations on a 32$\times$32$\times$1 supercell. The BKT phase was identified there as well, with its temperature region ranging between15.6 -- 17.5 K (that is rather close to aforementioned results using a 20$\times$20$\times$1 supercell).

As detailed in SM,  we also conducted additional simulations on relaxed 2UC and 3UC films made of SnTe. For these thicker films, no-intermediate BKT phase was detected in-between the ferroelectric and paraelectric states. Such thickness-induced disappearance of the BKT phase is perfectly in-line with the fact that such complex topological state is basically a 2D phenomenon and that out-of-plane interactions between layers tend to destroy it  \cite{berezinsky1972destruction,kosterlitz1973ordering}.

The effect of epitaxial strain on physical properties was also studied for the 1UC SnTe film, by imposing the following constraints on some elements (in Voigt notation) of the $\epsilon$ strain tensor: $\epsilon_1 = \epsilon_2 = \epsilon_{misfit}$, with $\epsilon_{misfit}$ ranging between -1\% and 5\%, while $\epsilon_6 = 0$ for any selected $\epsilon_{misfit}$ \cite{note1}. Note that the zero in $\epsilon_{misfit}$ corresponds to the lattice constant of the equilibrium paraelectric phase and that the `1' and `2' indices of $\epsilon_1$ and  $\epsilon_2$ refer to the a- and b-axis, respectively. For each $\epsilon_{misfit}$,  $T_{\rm C}$ is determined to the temperature at which the dielectric response peaks and $T_{\rm BKT}$ is identified to be
 the temperature at which the correlation function exponent $\eta$ is equal to 0.25.
The resulting temperature-{\it vs}-strain phase diagram, for $\epsilon_{misfit}$ varying between $\simeq$ -0.2\% and +0.8\%,   is displayed in Fig. 4.
$T_{\rm BKT}$ and $T_{\rm C}$ both increase linearly with increasing strain (from compressive to tensile). Particularly (see inset of Fig. 4), $T_{\rm C}$ reaches 300 K at $\epsilon_{misfit}$ = 3.3\% and goes up to 473 K for $\epsilon_{misfit}$ = 5\% -- which reveals the significant effects of strain on tuning $T_{\rm C}$. Such latter effects also result in $T_{\rm C}$ vanishing at  $\epsilon_{misfit}$ $\simeq$ -0.1\%, indicating the collapse of the FE phase for larger-in-magnitude compressive strain. Moreover, the BKT phase is found to exist in the strain window ranging from $\epsilon_{misfit}$ $\simeq$ -0.2\% to $\epsilon_{misfit}$ $\sim$0.8\%
(at which  $T_{\rm BKT} = T_{\rm C}=$ 71.4 K). Within this range, the maximal temperature interval at which this BKT phase is the equilibrium phase is equal  to 7 K and corresponds to  $\epsilon_{misfit}$ = -0.1\%. Note that for strains varying between -0.1\% and -0.2\%, we also numerically found, at low temperatures, that (i) the BKT phase has a very similar energy than the ideal PE phase (for which all local modes are zero); and (ii) dipole patterns can exhibit both closely-bound vortex-antivortex pairs (like  in the BKT phase) and ordered domains (like  in a FE phase), as shown in Fig. S4 of the SM. One can thus conclude that, within -0.1\% and -0.2\%, BKT phase can be stable and several low-energy metastable phases may coexist at low temperatures. 

In summary, a first-principle-based effective Hamiltonian was combined with  MC simulations to reveal many fingerprints of a BKT phase transition in {\it relaxed} 1UC  SnTe film, for  a temperature window bridging the
low-temperature FE and high-temperature PE phases. Moreover, epitaxial strain is demonstrated to affect the stability region of BKT phase, such as BKT phase can be stable at the lowest temperature within strain range from -0.1\% to -0.2\%.  Such predictions therefore render the class of materials made by 2D  ferroelectric monolayers as  new candidate systems to observe the BKT phase, as well as point out the importance of strain to control some aspects of such fascinating phase.
We hope that our findings deepen the current knowledge on topology in condensed matter, in general, and BKT physics, in particular.

{\it Note added:} During the review of our manuscript, we noticed a recent work that reports vortex-antivortex pairs in 1UC SnSe from {\it ab initio} molecular dynamics \cite{villanova2020theory}.

\begin{acknowledgments}
C.X. and L.B. thank Office of Basic Energy Sciences under contract ER-46612 and an Impact Grant from Arkansas Research Alliance. S.P., Y.N. and L.B. thank  the  financial support of the  DARPA  Grant  No.  HR0011727183-D18AP00010  (TEE  Program). Y.N.  and  L.B.  also  acknowledge  support  of  the  ARO  grant  W911NF-16-1-0227. H.X. is supported by NSFC (11825403), the Special Funds for Major State Basic Research (2015CB921700), Program for Professor of Special Appointment (Eastern Scholar), Qing Nian Ba Jian Program, and Fok Ying Tung Education Foundation. The Arkansas High Performance Computing Center (AHPCC) is also acknowledged.
\end{acknowledgments}


\begin{thebibliography}{43}
\bibitem [{not()}]{note1}%
  \BibitemOpen
\bibfield  {journal} {  }\href@noop {} {\bibinfo  {journal} {Note that, in the
  case of fully relaxed 1UC SnTe film (for which results are shown in Figs
  1-3), we numerically found that $\epsilon_1$ = 0.3\%, $\epsilon_2$ = -0.09\%
  and $\epsilon_6$ = -0.01\% at low temperatures, which yield $T_{\rm BKT}$ and
  $T_{\rm C}$ being similar to those of the epitaxial film associated with
  $\epsilon_{misfit}$ = 0.07\%}\ }\BibitemShut {NoStop}%
\end{thebibliography}

%

\end{document}